\documentclass[10pt,conference]{IEEEtran}
\pdfoutput=1
\usepackage{amsmath,amsfonts}
\usepackage{algorithmic}
\usepackage{algorithm}
\usepackage{array}
\usepackage{textcomp}
\usepackage{stfloats}
\usepackage{url}
\usepackage{verbatim}
\usepackage{pdfpages}

\usepackage{graphicx}
\usepackage[noadjust]{cite}
\usepackage[justification=centering]{caption}
\usepackage{tikz}
\usepackage{subcaption}
 \usepackage[normalem]{ulem}
\usetikzlibrary{spy}
\usepackage{amsmath}
\usepackage{amssymb}
\usepackage{tikz} % To generate the plot from csv
\usepackage{pgfplots}
\usepackage{tabularx}
\usepackage{xcolor}
\usepackage{verbatim}
\pgfplotsset{compat=newest}
\newcommand{\robbert}[1]{{\color{magenta}[Robbert: #1]}}

\newcommand{\C}{\mathbb{C}}
\newcommand{\E}[1]{\mathbb{E} \{ #1 \}}

\IEEEoverridecommandlockouts

\title{Cell-Free Massive MIMO in the O-RAN Architecture: Cluster and Handover Strategies}%
\author{\IEEEauthorblockN{Robbert Beerten, Vida Ranjbar, Andrea P. Guevara, Sofie Pollin}
\IEEEauthorblockA{\textit{Department of Electrical Engineering, KU Leuven, Belgium} \\
Corresponding Author: \{robbert.beerten\}@kuleuven.be}%
\thanks{This research has received funding from the European Union’s Horizon 2020 research and innovation programme under Grant Agreement No. 101017171 (MARSAL project).}}%}%
\begin{document}%
% \markboth{Journal on Selected Areas in Communications}%
% {Shell \MakeLowercase{\textit{et al.}}: A Sample Article Using IEEEtran.cls for IEEE Journals}%
\maketitle%
\begin{abstract}
Cell-free Massive MIMO has emerged as a promising solution for next-generation radio networks. Particularly the user-centric variant where users are served by a limited subset of access points, a so-called cluster, has garnered significant attention within the research community. Despite numerous proposed physical layer solutions, managing AP   
clusters in case of user mobility remains challenging.
In this study, we first present a realistic method for modeling the temporal evolution of the channel in cell-free Massive MIMO. Next, we develop two clustering and handover strategies: 1) a fixed clustering strategy that computes the ideal cluster when a cluster handover threshold is exceeded and 2) an opportunistic strategy that opportunistically adds serving APs 
as the user moves.
Moreover, we establish a connection between our findings and O-RAN, an emerging network architecture that offers open interfaces and network-wide control capabilities, thus, facilitating practical implementation of our research.
% Through our comprehensive investigation, we aim to unlock the potential of cell-free Massive MIMO, enabling seamless user experiences and empowering the next generation of wireless networks with unparalleled network control.
\end{abstract}

\section{Introduction}
% \vida{large paragraph...}
Cell-Free Massive MIMO (CF mMIMO) has emerged as a prominent enabler of next-generation networks, primarily due to its capacity to provide uniform service and exceptional coverage \cite{7421222}.
CF mMIMO refers to a network with many more Access Points (APs) than User Equipment (UEs), where APs cooperate via coherent joint transmission and reception.
Recently, user-centric CF mMIMO \cite{Buzzi_DAndrea_2017} has gained considerable traction as the most promising variant of this technology. In this case, users are served by a limited subset of APs, a so-called cluster, that provide the best channel. % \sofie{best channel or other metric?} 
However, efficiently managing these networks in the presence of user mobility remains a significant concern. 
While the problem was studied in \cite{DAndrea_Interdonato_Buzzi_2021,Zaher_Bjornson_Petrova_2022, Xiao_Mahonen_Simic_2022}, these works 1) lack a tractable channel model to study handover in sub-6GHz correlated Rayleigh fading 
and 2) rely on a single Central Processing Unit (CPU) for cluster reformulation 
and processing of the higher parts of the physical layer. 
Scalability of CF mMIMO networks was recently recognized as an important requirement \cite{Bjornson_Sanguinetti_2020}. The centralization of all computations in a single CPU should be avoided as 
this imposes immense requirements on the CPU's fronthaul capacity and computing power.
Hence, we provide a method where APs can locally decide to serve UEs,
thereby mitigating the computational burden on the CPU.

Splitting the CPU into multiple local computing units is also explored in \cite{Interdonato_Frenger_Larsson_2019, Li_Sun_Ji_Chen_2022, Riera-Palou_Femenias_2019}. % \sofie{Local estimates are then combined using a set of possible methods?}
Local signal estimates of these CPUs are then combined via different methods such as for example large-scale fading decoding.

This idea is reflected in our work, where multiple
processing units (O-DUs) decode the signal collaboratively. 
Recently, O-RAN came forth as a promising %novel 
approach to organise mobile networks. 
This new architecture proves to be interesting for practical
deployments of CF mMIMO networks for two reasons.
First, the physical layer is split between the O-RAN Distributed Units (O-DUs) 
and Radio Units (O-RUs) for the high PHY and low PHY, respectively.
This is highly parallel to seminal papers on CF mMIMO where physical layer processing is split between 
the APs and the CPU. % and thus very similar to the O-RU and O-DU, respectively.
Multiple cooperating O-DUs 
can then be linked to theoretical CF mMIMO literature considering multiple cooperating CPUs.
% (as a split 7.2x in 3GPP terms). 
Second, extra functional
blocks are introduced, enabling AI and containerized service orchestration on the network. 
These include the Near-Real Time RAN Intelligent Controller (Near-RT RIC), the Non-RT RIC, and 
the Service Management and Orchestration (SMO) framework. % \cite{ORAN-OVERVIEW}. 
% This new disaggregated and intelligent RAN architecture can enable O-RUs to serve UEs collaboratively. 
The functional split and options for network-wide control can achieve cooperation 
amongst O-RUs, even beyond the borders of the O-DUs.
% We see great value %\vida{s} 
% in these two concepts and especially in combining them. 
In this work,
we show considerable synergy between these two and demonstrate that their combination  
can enable the next generation of wireless networks. 
% \subsection{O-RAN}
% hence also requiring these O-DUs to cooperate.
% \sofie{I miss here a bit a statement on distributed O-RU that cooperate, before you make the jump to the O-DUs that cooperate....}
This work will focus on how an O-RAN network 
can achieve cooperation and smooth handover between its O-DUs
% \sofie{at this point I think: what is the link between O-DU and the clusters you mentioned before... ?}
via our proposed inter-O-DU interface, the E2 interface and the Near-RT RIC.
\paragraph*{Contribution}
% \ag{under O-RAN architecture. (In fact, for the rest of the paper you do not use AP/CPU anymore, better stick to O-RAN terminology since the beginning)}. 
This paper makes three key contributions to the current state-of-the-art. First, we propose a tractable temporal channel model for CF mMIMO at sub6-GHz frequencies. 
Second, we study the problem of handover in CF mMIMO and introduce two strategies: 
an opportunistic clustering strategy, where APs can locally decide to serve
specific users, % \sofie{and estimate their signals locally and send this estimate to a central controller?}
and a
fixed clustering strategy, where handover is computed % \vida{decided?} 
at a central processor. Our work establishes a framework that includes our notions of primary O-RU, primary O-DU and serving and measurement cluster.
Third, we extend our previous work \cite{selfCite} and map our handover solutions
to a real network architecture and offer practical
implementation guidelines. % \sofie{still not sure what an abstract CPU is}. 
Thus, this works bridges theory and practice by linking the algorithms to the O-RAN architecture.
% \robbert{add to contribution}
% 1) we frame handover in a O-RAN cel-free network as a definition of a primary O-RU, primary O-DU and serving and measurement cluster.
% 2) we compare two methods to evolve serving and measurement cluster: centralised hard reclustering versus opportunistic soft reclustering (just came up with hard and soft but maybe the terms are quite informative? as with hard you mean the cluster is recomputed totally)? - but maybe you can say 
% we compare two methods to evolve serving and measurement cluster: centralised fixed reclustering versus opportunistic reclustering
% 3) we evaluate the methods using a newly proposed spatially correlated channel model.
\paragraph*{Notation}
 A diagonal matrix with the elements $x_i$ on the main diagonal is denoted by $\text{diag}(x_0, x_1, \cdots, x_N)$.
 The cardinality of a set $\mathcal{S}$ is written as $|\mathcal{S}|$. For matrices and vectors alike, $\mathbf{A}^T$ is the transpose,
 $\mathbf{A}^H$ is the Hermitian conjugate of the matrix, and $\mathbf{A}^{\ast}$ is the complex conjugate. 
 The $l$'th element of vector $\mathbf{a}$ is denoted by $[\mathbf{a}]_l$. The expected value of $x$ is denoted by 
 $\E{x}$. 
 
% We also list our commonly used symbols 
%  in Table~\ref{table:symbols}.

% \input{symbols.tex}
%%%%%%%%%%%%%%%%%%%%%%%%%%%%%%%%%%
\section{Cluster Formation } % \sofie{and Reformation} \robbert{I have a separate section for handover?}} 
%%%%%%%%%%%%%%%%%%%%%%%%%%%%%%%%%%
\label{sec:cluster}
In this work, we consider a scenario with $K$ UEs, $C$ O-DUs, $L$ O-RUs with $N$ antennas each 
and a single Near-RT RIC. We also denote the set of O-RUs that are connected to O-DU $c$ as $\mathcal{L}_c$.
When a UE first connects to the network, it selects a \textbf{primary O-RU}, and the O-DU that controls that O-RU logically becomes the primary O-DU. This choice is based on the channel gain that the UE computes based on a DL pilot\footnote{We see this as a valid assumption as many works have argued in favour of DL pilots for DL 
signal processing, which is out of scope for this paper yet important for CF mMIMO systems \cite{Interdonato_Ngo_Frenger_Larsson_2019}.}.
 The concept of the primary O-RU is important as it determines \textbf{handover} and the
 \textbf{measurement cluster}. We define handover as a transition to a new
 primary O-RU.
 The measurement cluster, $\mathcal{M}_k^m$, consists of the O-RUs closest to that UE's primary O-RU on which the channel gain for a specific UE $k$ is measured. 
 Subsequently, the \textbf{serving cluster} % \sofie{I would put primary O-RU and serving cluster also in bold first time they are used} 
 $\mathcal{M}^{\text{s}}_k$
are formed as a subset of $\mathcal{M}_k^m$ by using one of the two cluster formation algorithms explained in the following subsections.
This serving cluster performs joint precoding/decoding for a specific UE $k$.
Note that these O-RUs in $ \mathcal{M}^{\text{s}}_k$ can belong to one or multiple O-DUs, thus, potentially requiring O-DUs to cooperate. We also define $\mathcal{D}_l$ as the set of all UEs that are served by O-RU $l$. 
Two strategies for determining {\bf serving clusters}  are proposed: a \emph{fixed} serving cluster formation and an \emph{opportunistic} serving cluster formation. 
We compare these two methods against two baselines, ubiquitous CF and a cellular system with a single O-DU serving. 
%%%%%%%%%%%%%%%%%%%%%%%%%%%%%%%%%%%%%%%%%%%%%%%%%%%%%%%%%
\subsection{Fixed Cluster Formation}%
%%%%%%%%%%%%%%%%%%%%%%%%%%%%%%%%%%%%%%%%%%%%%%%%%%%%%%%%%
\label{subsec:uetrig}
After the primary O-RU, O-DU and measurement cluster are determined, 
the O-DUs that control O-RUs in the 
measurement cluster report the UL channel gains, $\beta_{l,k}$, between those O-RUs $l$ and %that 
UE $k$ to the Near-RT RIC. 
To determine the serving cluster, we define $O_k[t]$ as the function that selects the subset of $|\mathcal{M}^{\text{s}}_k|$ O-RUs in the measurement cluster 
that experience the highest UL channel gains from UE $k$ at time $t$.
%The primary O-DU requests the Near-RT RIC to calculate a serving cluster based
%on the channel gain measurements from the O-RUs. 
% In Algorithm~\ref{alg:UCInitial}
% we show how the fixed serving cluster is formed for a specific UE $k$ by the Near-RT RIC. 
These serving O-RUs then become the serving cluster. Next, the sum of DL channel gains from the initial serving cluster to UE $k$, is determined as $\bar{P}_k$ at the UE. 
% \ag{I am a bit confused here, the channel gain is similar to the pathloss, why it is different for UL and DL?}\vida{yes why we changed from UL to DL here?}\robbert{In this method the UE triggers handover, the decision is based on the total DL received power. This way you can do it without communication between O-DUs}. 
This metric represents the total signal quality in the cluster when it is created 
and plays a crucial role in the handover mechanism, which will be discussed in Section~\ref{sec:handover}.
 The main advantage of this method is that the number of O-RUs, $|\mathcal{M}^{\text{s}}_k|$, 
can be %chosen 
controlled by the Near-RT RIC. The main disadvantage is the substantial signaling between the O-DUs and the Near-RT RIC. %if the channel gains $\beta_{l,k}$ are communicated frequently}. 
\subsection{Opportunistic Cluster Formation}
\label{subsec:nettrig}
By the opportunistic method, after the allocation of the primary O-RU and O-DU,
% and measurement cluster
O-DUs can decide autonomously if they
serve a specific UE on O-RUs with 
available resources without involving the Near-RT RIC. In this case, the Near-RT RIC serves 
as a bookkeeper and to inform the O-DUs about two things.
First, O-DUs must be aware of which O-DUs serve as primary O-DU for UEs 
they serve themselves. Second, they must know which UEs have its O-RUs in their measurement cluster. Both of these only
change in the case of primary O-RU changes.
These message are necessary such that an O-DU knows with which O-DU it should instantiate an
inter-O-DU interface for cooperative decoding and to limit the number of UEs an O-DU keeps 
track of.
Thus, this method has very limited requirements in terms of signaling. O-DUs decide opportunistically by only selecting O-RU - UE pairs with the highest UL channel gains.
The O-DUs prioritise pairs of UEs with primary O-RUs rather than opportunistic serving.
\cite{Buzzi_DAndrea_2017} also proposes a clustering scheme  
where O-RUs can decide locally if they serve a specific UE but do not 
explore how to change these UE - O-RU associations in case of UE mobility. %\sofie{Can you maybe also explain how it is different if the O-RU or the O-DU decides? }.
% O-DUs decide locally by measuring which UEs have the highest UL channel gains. 
To limit the size of the serving cluster, we restrict
the size of the measurement cluster
and also limit the number of UEs a specific O-RU can serve to $N$. 
In contrast
to the fixed clustering strategy, this opportunistic strategy avoids
%he triggering of 
a network-wide procedure executed at the Near-RT RIC.
The function  $Q^{(w)}_l[t]$ maps all UEs whose measurement cluster 
contains O-RU $l$ to the subset of $w\leq N$ UEs that achieve the best channel towards O-RU $l$ at time 
$t$.
% \vida{I would explain the algorithm better. I mean I dont know why you need $Q_l^{(N - K^{\star}_l)}[0]$} \\
This function %is needed to ensure that 
loads an O-RU %can be loaded 
up to its limit, including the UEs that use it as 
primary O-RU i.e. an O-RU that currently serves $K^{\star}_l$ UEs as primary O-RU could take
an additional $N - K^{\star}_l$ UEs opportunistically.
% \sofie{very confusing...  is the max only for those O-RU that are also a primary O-RU? how does this link to $w$ from before... }. 
Algorithm~\ref{algo:NCInitial} shows how O-RUs can be maximally loaded.
% Notice the importance of the measurement cluster to the number of O-RUs that serve 
% a specific UE, as only O-RUs in its measurement cluster can potentially be used to serve that UE. 
% \cite{Bursalioglu_Caire_Mungara_Papadopoulos_Wang_2019} 
% The goal is ensuring that an O-RU is not serving more UE than the number of antennas $N$, which means that we can assign each O-RU $l$ 
% at max $N-K^{\star}_l$ UE, where $K^{\star}_l$ is the number of UE the O-RU is already serving. 
The main advantage of this method is the low computational cost and signaling load. The main disadvantage of this method is that there are no guarantees 
on the number of serving O-RUs beyond the primary O-RU. 
\begin{algorithm}[hb]
\caption{Initial Opportunistic Cluster Formation % \sofie{at time $t=0$}
}\label{algo:NCInitial}
\begin{algorithmic}[1]
\FOR{$k = 1 \dots K$} 
    \STATE{\textbf{Assign primary O-RUs}}
    \STATE $l^{\ast}_k = \arg\max_l \beta_{l,k}[0]$
    %\STATE{$\bar{\beta}^{\text{dB}}_{l^{\ast}_k, k} \gets \beta^{\text{dB}}_{l^{\ast}_k, k}[0]$}
\ENDFOR
\FOR{$l = 1 \dots L$} 
      \STATE{\textbf{Load each O-RU upto $N$ UEs}}
      \STATE $\mathcal{D}_l[0] \gets Q_l^{(N - K^{\star}_l)}[0]$  
%      \STATE{$\bar{\beta}^{\text{dB}}_{l, k} \gets \beta^{\text{dB}}_{l, k}[0]  \qquad \forall k \in \mathcal{D}_l$}
\ENDFOR
\end{algorithmic}
\end{algorithm}
\subsection{Baselines}
By the ubiquitous method,
% \ag{, also known as canonical cell-free system}
every UE is served by every single O-RU in the network and thus $|\mathcal{M}^s_k| = L$. 
% \ag{You do not define what it is \textit{L} yet.} 
We use this as an upper bound on the performance of the clustering.
% It is expected that performance %is diminished
% drops when 
% only part of the network is used to serve a specific UE. 
By comparing to the optimal, ubiquitous case, we can evaluate
the performance gap to a heuristic handover scheme. 
For the cellular case, every UE is served only by O-RUs connected to a single O-DU. 
% Hence, only $|\mathcal{M}^s_k| = L/C$ O-RUs cooperate to serve a single UE. 
This is used as a lower bound on the performance to show the 
benefits of implementing our proposed inter-O-DU interface. 
The disadvantages are two-fold; first the amount of O-RUs is inherently limited due to the size of 
the O-DU; second, the set is rarely optimal due to the high probability of O-RUs from a different 
O-DU providing a better channel than the selected O-RUs. 
% This method of operation is similar to canonical Distributed MIMO.
\section{System Model}
\label{section:system_model}
% \sofie{I wonder if it would not be better to put system model before the cluster section...}
%currently present in the  state-of-the-art 
This section describes %how to tackle 
UL detection, channel estimation, and combiner calculation 
during a specific coherence 
block. We use a method from the current state-of-the-art and map it to the 
O-RAN architecture.  
Subsections \ref{subsection:system}, \ref{subsection:chest}, and \ref{subsection:combining} %is
use the method proposed in \cite[Ch. 4-5]{SIG-109}.
In Section~\ref{section:mobility}, we extend our system to allow for mobility of the 
UEs and how to temporally evolve our channel model. 
% Section~\ref{sec:handover} 
% then discusses how the clusters should be reformed based on the 
% mobility of the UEs. 
% Finally, we provide practical guidelines on how to integrate our solutions into the O-RAN framework 
% and specifically the Near-RT RIC in Section~\ref{section:oran}.
% \sofie{You say 'this method' a lot, but I still have not clue what you refer to. Also, it seems a bit late to only map it in Section 6? What will do you in between? Can you make things more explicit for the reader. E.g., You detail the system model without mobility in this Section, IV will then extend the system model with mobility, including a channel model. Section V then details how to do handover in case of user mobility, and what it means for primary and serving O-RU, en then all is mapped to O-RAN in Section VI. The reader should be informed of this all I think. Also, you might want to add the mobility section to your system model. }. 
%%%%%%%%%%%%%%%%%%%%%%%%%%%%%%%%%%
\subsection{Channel Model}
%%%%%%%%%%%%%%%%%%%%%%%%%%%%%%%%%%
\label{subsection:system}
In this work, we consider UL detection. 
% We assume each O-RU to have $N$ antennas.
The channel vector between a single-antenna 
UE $k$ and an O-RU $l$, $\mathbf{h}_{l,k} \in \C ^{N}$,
is generated from a complex normal distribution,
\begin{equation}
    \mathbf{h}_{l,k} \sim \mathcal{CN}(\mathbf{0}, \mathbf{R}_{l,k}).
\end{equation}
%Where
Where $\mathbf{R}_{l,k} \in \C^{N \times N}$ is the covariance matrix of the channel. 
% This model implies that we assume the channels between one UE and different O-RUs to be 
% mutually uncorrelated. This is a valid assumption if the O-RUs are placed many wavelengths apart. 
We assume these $\mathbf{R}_{l,k}$ to be known at the respective O-DU for that O-RU $l$.
% Some methods exist to estimate these from a relatively limited amount of samples \cite{Ranjbar_Moonen_Pollin_2021}. 
We will later link these $\mathbf{R}_{l,k}$ to our mobility model in Section~\ref{section:mobility}.
%%%%%%%%%%%%%%%%%%%%%%%%%%%%%%%%%%
\subsection{Channel Estimation}
%%%%%%%%%%%%%%%%%%%%%%%%%%%%%%%%%%
\label{subsection:chest}
To estimate the channel, the UEs transmit mutually orthogonal pilot sequences $\boldsymbol{\phi}_i \in \mathbb{C}^{\tau_p}$. 
% The O-DU can exploit this orthogonality to separate the UEs while estimating the channel. 
\begin{comment}
\begin{equation}
    \label{eq:ortho}
      \boldsymbol{\phi}_i^H\boldsymbol{\phi}_j = 
        \begin{cases}
        \tau_p & \text{if } j = i \\
        0 & \text{if } j \neq i    
        \end{cases}.
\end{equation}
\end{comment}
In the UL, UE $k$ transmits  %ir
pilot $\boldsymbol{\phi}_{t_k}$ %\sofie{bold? }% 
in the designated time slot, where $t_k$ is 
the index of the pilot allocated to UE $k$. 
%The O-RUs receive %r 
O-RU $l$ receives
the superposition of all of the UEs' pilots %is 
as $\mathbf{Y}_{l} \in \mathbb{C}^{N \times \tau_p}$:
\begin{equation}
    \label{eq:decorr}
      \mathbf{Y}_{l} = \sum_{k=1}^{K} \sqrt{p_k} \mathbf{h}_{l,k} \boldsymbol{\phi}_{t_k}^T + \mathbf{N}_{l} .
\end{equation}
Where $\mathbf{N}_l$ is a matrix with the measured noise realizations $\mathbf{n}_{t,l}$ in the columns. Each of 
its elements are i.i.d. distributed white noise from the distribution $\mathcal{CN}(0, \sigma^2_{\text{ul}})$.
% The transmit power of UE $k$ is $p_k$. The O-DU can estimate the respective channels per O-RU as these O-RUs are assumed to be many wavelengths apart and thus the 
% channels between a specific UE and 
% multiple O-RUs are mutually uncorrelated. 
The O-DU decorrelates these received pilots for UE $k$ in O-RU $l$ as $\mathbf{y}_{l,k}^{(\text{p})} = \mathbf{Y}_{l}\frac{\boldsymbol{\phi}_{t_k}^{\ast}}{\sqrt{\tau_p}} $. 
\begin{comment}
\begin{equation}
     \mathbf{y}_{l,k}^{(\text{p})} = \sum_{i=1}^{K} \sqrt{\frac{p_i}{\tau_p}} \mathbf{h}_{l,i} \boldsymbol{\phi}_{t_i}^T \boldsymbol{\phi}_{t_k}^{\ast} + \mathbf{N}_{l} \boldsymbol{\phi}_{t_k}^{\ast}. 
\end{equation}
\end{comment}
% \sofie{so you miss a $\sqrt{\tau_p}$? }
% Afterwards, only interference from UEs that share the same pilot remains, which is
% known as pilot contamination \cite{Jose_Ashikhmin_Marzetta_Vishwanath_2011}. 
% UEs sharing their pilot with UE $k$ are denoted as the set $\mathcal{P}_k$,
% \begin{equation}
%       \mathbf{y}_{l,k}^{(\text{p})} = \sum_{i \in \mathcal{P}_k} \sqrt{p_i \tau_p} \mathbf{h}_{l,i} +  \mathbf{N}_{l} \boldsymbol{\phi}_{t_k}^{\ast}.
% \end{equation}
% \sofie{Did not check the details, but should it maybe be $\sqrt{p_i \tau_p }$ }
The O-DU then calculates the MMSE channel estimates between UE $k$ and O-RU $l$, $\hat{\mathbf{h}}_{l,k}$, as 
\begin{equation}
  \hat{\mathbf{h}}_{l,k} = \sqrt{\tau_p p_k} \mathbf{R}_{l,k} \left(  \sum_{i \in \mathcal{P}_k} \tau_p p_i \mathbf{R}_{l,i} + \sigma_{\text{ul}}^2 \mathbf{I}_N \right)^{-1}   \mathbf{y}_{l,k}^{(\text{p})}.
\end{equation}
We denote the error on the channel estimate as 
   $ \tilde{\mathbf{h}}_{l,k} = \hat{\mathbf{h}}_{l,k} - \mathbf{h}_{l,k}$.
The covariance of the channel estimation error is $\mathbf{C}_{l,k}$.
% \sofie{why $i$?}
% For this work, we do not target the pilot assignment but we see it as a logical extension to 
% this work. We also firmly believe that it makes much sense to execute the pilot assignment 
% at the Near-RT RIC since this component has an almost network-wide view. 
% Adressed in the results section already

% \sofie{Probably this section on the channel estimation can be compressed. You mainly need to know that the O-DU does it.}
% \sofie{Do we also need to explain beta?}
%%%%%%%%%%%%%%%%%%%%%%%%%%%%%%%%%%
\subsection{Receive Combining}
%%%%%%%%%%%%%%%%%%%%%%%%%%%%%%%%%%
\label{subsection:combining}
% \robbert{This should be decreased to half a column}
Large-Scale Fading Decoding (LSFD) %\cite{SIG-109} 
fits nicely into the O-RAN architecture. 
% We employ this decoding scheme by having 
% the primary O-DU calculate LSFD weights.
UL detection has three stages in our system. 
First, each O-RU $l$ estimates the signal $\hat{s}_{l,k}$ of its served UEs locally and sends 
these local estimates to its respective O-DU. Second, this O-DU $c$ computes more accurate 
estimates for the UE's signal $\hat{s}_{k}^{c}$ based on its LSFD weights 
% \vida{why did you repeat $\hat{s}_{k}^{c}$ here?} $\hat{s}_{k}^{c}$ 
which the primary O-DU then combines into the final estimate for the UE's signal, $\hat{s}_k$.
This mapping of LSFD to O-RAN was proposed in \cite{selfCite} and relies on the interface 
between cooperating O-DUs which was proposed in \cite{Ranjbar_Girycki_Rahman_Pollin_Moonen_Vinogradov_2022}. 
The received signal at each O-RU is, 
\begin{equation}
    \mathbf{y}_l = \sum_{k=1}^{K}  \mathbf{h}_{l,k} s_k + \mathbf{n}_l .
\end{equation}
In our %example 
analysis,
we use the Local Partial MMSE (LP-MMSE) combiner
% \cite{Bjornson_Sanguinetti_2020}
as it is computed separately for each O-RU by the relevant O-DU. % \cite{ORAN-FH}.   
% We define $\mathcal{D}_l$ as the set of all UEs that are served by O-RU $l$. 
% The fronthaul specification in O-RAN \cite{ORAN-FH} supports both transfer of the receive combining vector 
% from the O-DU to the O-RU over the fronthaul as well as transfer of CSI to the O-RU such that the 
% O-RU can calculate the combining vector itself. Here, we assume the first option.
% \sofie{So, why can't you then do some kind of local receive combining vector computation at the O-DU level? over the set of UE that are served by the O-DU?  maybe too complex would be some kind of multi-cell MMSE - and you assume no correlation anyway across O-RU} 
% \robbert{This way the O-DU only has to perform channel estimation for specific UEs on O-RUs it is using 
% to serve those UEs}. 
The O-DU computes an LP-MMSE receive combiner $ \mathbf{v}_{l, k}$ for each of its O-RU - UE associations,  
\begin{equation}
    \label{eq:lpmmse}
    \mathbf{v}_{l, k} = p_k 
    \left(\sum_{i \in \mathcal{D}_l} p_i \left(\hat{\mathbf{h}}_{l,i}\hat{\mathbf{h}}_{l,i}^{H} + \mathbf{C}_{l,i}\right) + \sigma_{\text{ul}}^2\mathbf{I}_N \right)^{-1} \hat{\mathbf{h}}_{l,k} .
\end{equation}
 The O-DU computes
this combiner only if $k \in \mathcal{D}_l$.
% \vida{I now realized that you defined $\mathcal{D}_l$ only in opportunistic cluster formation and not generally (also for fixed cluster formation). Maybe you can define $\mathcal{D}_l$ where you defined measurement and serving clusters?}.
Otherwise 
we take $\mathbf{v}_{l,k} = \mathbf{0}_{N \times 1}$.
The O-RU then locally combines the received signal via $\mathbf{v}_{l, k}$ as,
\begin{equation}
    \hat{s}_{l,k} = \mathbf{v}_{l,k}^H \mathbf{y}_{l,k}, \quad \forall l \in \mathcal{M}^{\text{s}}_k .
\end{equation}
We compute cluster-wide LSFD combining weights for the local O-RU estimates.
$\mathbf{g}_{ki}$ is the effective gain for UE $i$ in the receive combiner for UE $k$ and $\delta_{l,k}$ indicates if UE $k$ is served by O-RU $l$ i.e.
% $
%     \mathbf{g}_{ki} = \begin{bmatrix} 
%         \delta_{1,k}\mathbf{v}_{1, k}^H\mathbf{h}_{1,i}  &
%         \delta_{2,k}\mathbf{v}_{2, k}^H\mathbf{h}_{2,i}  &
%         \dots &
%          \delta_{L,k} \mathbf{v}_{L, k}^H\mathbf{h}_{L,i} 
%     \end{bmatrix}^T .
% $
$
    \mathbf{g}_{ki} = \begin{bmatrix} 
        \mathbf{v}_{1, k}^H\mathbf{h}_{1,i}  &
        \mathbf{v}_{2, k}^H\mathbf{h}_{2,i}  &
        \dots &
        \mathbf{v}_{L, k}^H\mathbf{h}_{L,i} 
    \end{bmatrix}^T .
$
We denote the set of UEs that share at least one O-RU with UE $k$ as $\mathcal{S}_k$.
We use so-called n-opt LSFD \cite{SIG-109}, where it is assumed that only those UEs in $\mathcal{S}_k$
% \vida{How do you select the users in $\mathcal{S}_k$? Also check out computational cost section} 
induce significant
interference for UE $k$, and thus only those UEs are considered in 
the LSFD combiner. 
% \sofie{In practice, this means that every O-RU should share those gains for all it's UE, with the primary O-DU of each UE it is serving. }
The primary O-DU can calculate the LSFD weights $\mathbf{a}_k \in \mathbb{C}^{L}$ as \cite{SIG-109},
\begin{equation}
\label{eq:lsfd}
    \mathbf{a}_{k} = p_k \left( \sum_{i \in \mathcal{S}_k} \E{\mathbf{g}_{ki}\mathbf{g}_{ki}^H} + \mathbf{F}_k %+ \tilde{\mathbf{D}}_k 
    \right)^{-1} \E{\mathbf{g}_{kk}}.
\end{equation}
Where 
% $
%     \mathbf{F}_k = \sigma_{\text{ul}}^2 \text{diag}\left(\E{\| \delta_{1, k} 
%     \mathbf{v}_{1,k}\|}, \cdots, 
%     \E{\| \delta_{L, k}\mathbf{v}_{L,k}\|}\right).
% $
$
    \mathbf{F}_k = \sigma_{\text{ul}}^2 \text{diag}\left(\E{\|  
    \mathbf{v}_{1,k}\|}, \cdots, 
    \E{\|\mathbf{v}_{L,k}\|}\right).
$
The weight vector for UE $k$, $\mathbf{a}_k$, has $|\mathcal{M}_k^{\text{s}}|$ non-zero elements, one for each O-RU that is serving UE $k$. 
The LSFD weights are only a function of the statistics of the channels and channels 
% \vida{I see statistics as variance, mean, so maybe better to say channel between one user and multiple O-RU are uncorrelated? even independent, based on emil's paper on scalable cell-free for example} of different O-RUs 
are independent, so it is sufficient to compute the local mean gains for each O-RU to determine the overall weights \cite[p.323]{SIG-109}.
This decreases the inter-O-DU signalling for computing the LSFD combiner 
significantly. Because, instead of continuously sending effective gains, only the expected
values of these effective gains must be transmitted.
%  The secondary O-DUs provide timely updates of these channel statistics to the primary O-DU. 
We consider the statistics on these effective gains to be known at the O-DU for each UE that is served by
its' O-RUs, these are then transmitted to the relevant primary O-DUs over our inter-O-DU interface. % \vida{serving cluster is for UE, here you imply that serving cluster is for O-RU} 
O-DU $c$ combines the signals locally for each UE served by O-RU $l \in \mathcal{L}_c$ as 
\begin{equation}
    \hat{s}_k^{c} =  \sum_{l \in \mathcal{L}_c} [\mathbf{a}^{\ast}_{k}]_{l} \hat{s}_{l,k} .
\end{equation}
% \sofie{Why are there square brackets around a? }\robbert{You need the $l$'th element of $\mathbf{a}_k$}
% The communication of these $\mathbf{a}_k$ between O-DU is an extension from \cite{selfCite}, we see it as an important part of this inter-O-DU interface.
% \vida{secondary O-DUs share their local $\hat{s}_k^c$ with primary O-DU, I assume}
The primary O-DU then computes the final estimate for the UE's signal by combining the estimates of each of the serving O-DUs, $\hat{s}_k^c$ as 
$    \label{eq:9}
    \hat{s}_k = \sum_{c} \hat{s}_k^{c} = \sum_{l \in \mathcal{M}_k^{\text{s}}}   [\mathbf{a}^{\ast}_{k}]_{l} \hat{s}_{l,k}$.
% The result from \eqref{eq:9} is the same solution as if a centralized CPU would compute the result via LSFD. 
% After all, it is the primary O-DU that will do an integrity check after the final 
% decoding and can thus base its recalculation on this check \sofie{what check?}. 
% We leave this recalculation of $\mathbf{a}_k$ to future work.
% \sofie{? is this not the key of handover/clustering?}\robbert{I do handover on the same timescale
% as calculation of the weights}.
% Due to how the calculations are organised, the result is transparent 
The result is then transparent to how the O-RUs are organised amongst the different O-DUs.
% \ag{Do you think the signalling figure can fit here?}
% ; 
% after all, 
% the n-opt LSFD weights
% are calculated per O-RU.
% In our proposed architecture, computations are divided amongst 
% the primary O-DUs; thus enabling us to scale the algorithm across
% a large number of O-RUs and O-DUs. 
%%%%%%%%%%%%%%%%%%%%%%%%%%%%%%%%
% \subsection{Spectral Efficiency}
%%%%%%%%%%%%%%%%%%%%%%%%%%%%%%%%
The SINR is estimated as 
% \vida{This is use-and-then-forget-bound?} \robbert{Yes for UL LSFD}
\cite[Eq. 5.26]{Demir_Bjornson_Sanguinetti_2021},
    \begin{equation}
        \label{eq:sinr}
        \begin{aligned}
            \gamma_k^{\text{ul}} = \frac{p_k |\mathbf{a}_k^H \E{\mathbf{g}_{kk}}|^2}
            {\mathbf{a}_k^H 
            \left( \sum_{i \in \mathcal{S}_k}p_i \E{\mathbf{g}_{ki}\mathbf{g}_{ki}^H} - p_k \E{\mathbf{g}_{kk}}\E{\mathbf{g}_{kk}^H} + \mathbf{F}_k\right)\mathbf{a}_k}
        \end{aligned}
    \end{equation}
The SE is then estimated as:
    $\text{SE}^{\text{ul}}_k = \log_2(1 +  \gamma_k^{\text{ul}})$,
which is used in Section \ref{sec:results} for performance evaluation.

\subsection{Proposed Temporal Channel Model}
\label{section:mobility}
Next, we explain how the channel model can be evolved temporally. 
The UEs move in a straight line at constant speed $v_k$ at a random angle. % $\theta_k \sim \mathcal{U}[0, 2\pi]$.
% The moving direction, $\theta_k$, and the 
% speed of the UE, $v_k$, are fixed for each simulation run. 
The sample time of the simulation is $T_{S}$.
% A UEs position at time $t$ is then,
% \begin{equation}
% \begin{aligned}
%     x[t] = x[t-1] + v_k T_{S} \cos{(\theta_k)} \\
%     y[t] = y[t-1] + v_k T_{S} \sin{(\theta_k)}
% \end{aligned}.
% \end{equation}
If the deployment of O-RUs is dense, a minor movement of a UE induces a significant change in the Angle 
of Arrival (AoA) for the received signal at an O-RU and thus changes the covariance matrix of 
the channel significantly (see Eq.~\ref{eq:covariance}). 
In \cite{DAndrea_Interdonato_Buzzi_2021}, individual NLOS paths are updated via Jakes autocorrelation model.
The autocorrelation coefficient is computed with a zero'th order Bessel J function as follows
\begin{equation}
    \rho = J_0 ( \pi D_s T_s).
\end{equation}
Where $D_s = 2f_c \frac{v}{c}$ is the doppler spread of the received signal. This function tends to zero even for relatively small $T_s$ in the context of cluster formation. Hence, this model 
generates almost purely random channels.
We propose a model more suitable for larger sampling times where even the NLoS paths are not strongly correlated. % \vida{I don't know how you conclude that the channel is not correlated...} 
We expect the large-scale fading to be correlated. Hence,
we model shadow fading as an autoregressive function.
In \cite{Zhenyu_Wang_Tameh_Nix_2008}, the 
spatial auto-correlation function for shadow fading is formulated as $e^{-\alpha d}$, where $\alpha$ is the 
reciprocal of the decorrelation distance and 
 $d$ the distance between the two considered points.
% is correlated as \vida{What does it mean to be correlated as $e^{-\alpha d}$, you mean the expected value of their multiplication? Maybe better to use the same notation as in the reference, it is written }
We take $\alpha$ to be $(20 \text{m})^{-1}$ which is the recommended value for a typical European city \cite{Zhenyu_Wang_Tameh_Nix_2008}.
% \cite{He_Zhong_Ai_Oestges_2015} indicates that the decorrelation 
% distance becomes larger for more rural areas. We consider a relatively dense deployment of O-RUs, which 
% is typical for an urban area, and thus our value for $\alpha$ is valid.
% We assume that the shadow fading follows the same distribution during our simulation.  
A UE moves a distance of $ v_k T_{S}$ between
two subsequent samples. We model the shadow fading $ F_{l,k}[t]$ at time $t$ as,
\begin{equation}
     F_{l,k}[t] = \rho_k F_{l,k}[t-1] + \sqrt{1 - \rho_k^2} F_{l,k}^{\text{new}}.
\end{equation}
Where $\rho_k = e^{-\alpha v_k T_{S}}$ is the correlation coefficient for two subsequent realizations, 
$F_{l,k}^{\text{new}}$ is a newly drawn shadow fading from the distribution $\mathcal{N}(0, \sigma_{\text{sf}}^2)$.
% Our model allows to have correlated shadow fading for different samples of a specific UE but not between different UEs.
% In the end, this leads to a path loss model which is constructed as
The large scale fading is then modelled as, 
%\sofie{it might be interesting to see the impact of more/less correlation on the SE and handover performance.  Just to quantify the impact of the model on the overall conclusions. } 
%\robbert{Sofie, I agree but it requires generating new datasets which takes quite some time}
\begin{equation}
    \beta^{\text{dB}}_{l,k}[t] = -34 - 38\log_{10}(d_{l,k}[t]) + F_{l,k}[t].
\end{equation}
Where 
% $a$ is the path loss at the reference distance, $b$ is the distance dependant path 
% loss coefficient, 
$d_{l,k}$ is the distance between a specific UE $k$ and O-RU $l$ expressed in meters. 
%This temporal evolution of the channel model is novel in the literature. 
% \vida{Vague sentence, why large sampling time make the NLOS uncorrelated?}. In this work we update the shadow fading coefficient as a whole and evolve it using an empirically validated model. 
% Shadow fading plays an important 
% role in the proposed solution because the measurement cluster is constructed by taking the closest O-RUs to the primary one.
% Since shadow fading occurs, the O-RUs that are closer to the UE might not be optimal as they could be (partially) blocked. O-RUs that 
% are further away might provide a higher channel gain. Generally, the probability of further O-RUs being better than 
% closer ones %rises with a rising 
% increases with increasing %variance on the 
% shadow fading variance. Although we have chosen a fixed measurement cluster, 
% it might be valuable to make this size dependent on shadow fading variance. 
% % \vida{I don't understand what you mean by optimal cluster of O-RUs. Isnot optimal cluster cluster of all O-RUs?}. 
% %\robbert{I reworked this part now} \\
% Because we consider a dense deployment of O-RUs, a relatively small movement of the UE might induce a big change in the angle of arrival for a particular UE - O-RU association\vida{you already said that}. 
The channel covariance matrices are regenerated at every new time instance.
To calculate the second-order statistics of the channel, we use 
the one-ring scattering model \cite{massivemimobook} to calculate the element at row $m$ and column $n$ of the covariance 
matrix,
\begin{equation}
    \label{eq:covariance}
    [\mathbf{R}_{l,k}]_{m, n}[t] = \beta_{l, k}[t] \int e^{2 \pi j d_H (n - m) \sin(\bar{\phi}_{l,k})}f(\bar{\phi}_{l,k},t)d\bar{\phi}_{l,k},
\end{equation}
% \vida{subscript n and m means row n and column n?}
where $f(\bar{\phi}_{l,k},t)$ is the time-dependent distribution of the possible angles of arrival between user k and O-RU $l$. 
% This distribution can be either Gaussian, Laplacian or uniform around the true angle of arrival 
% for the signal. 
For this work, we consider a uniform distribution and thus, 
% \begin{equation}
   $\bar{\phi}_{l,k} = \phi_{l,k} + \delta, \delta \sim U[-\xi, \xi],$
% \end{equation}
where $\phi_{l,k}$ is the true angle of arrival, and $\xi$ models the richness of the scattering environment.

\section{Handover}
\label{sec:handover}

We consider a network where a UE is served by only  %part of the total amount of 
a subset of the O-RUs in the network and is moving. %\ag{at speed \textit{v}}.
Thus, any selected subset of O-RUs will become suboptimal as the UE moves away from its cluster. 
% \sofie{The serving cluster needs to be updated, and at some point a new primary O-RU should be selected. The latter is referred to as a handover.  }
Hence, in this section, we propose updating strategies for the clusters from Section~\ref{sec:cluster}.
% For the fixed clustering in Section~\ref{subsec:uetrig}, 
% we define a threshold for the entire cluster based on the DL power received at the UE. 
% For the opportunistic cluster in Section~\ref{subsec:nettrig}, 
% we define a threshold per O-RU such that UEs select their primary O-RU and 
% the O-DUs can decide locally which UEs to serve on which O-RUs opportunistically. 
We discuss 
these handover procedures at discrete times $t$, which are spaced apart at the same intervals, $T_s$.
% Firstly whenever a handover of primary O-RU is triggered, \sofie{incomplete sentence?}
When the UE triggers a \textit{primary O-RU handover}, it selects a new primary O-RU, and thus O-DU, with the highest DL channel gain.
% \vida{Actually selecting a primary O-RU is same as in cluster formation section. why there you did not mention primary O-RU is selected based on DL pilot and here only mention that primary O-RU is reselected?}.
The corresponding primary O-DU then requests a new
measurement cluster around the new primary O-RU. 
Every time we sample the clusters, % \ag{coherence time?}
we first check if a new cluster is needed, possibly update the cluster 
and then always update the LSFD weights.
\subsection{Handover for Fixed Clustering}
% \robbert{Removed the pseudocode, still need to write in text}
% \begin{algorithm}[h!]
% \caption{Handover for Fixed Clustering }\label{alg:cap}
% \begin{algorithmic}[1]
%  \STATE \textbf{Initialize:}
%  \STATE \hspace*{\algorithmicindent}\parbox[t]{.8\linewidth}{\raggedright  $\mathcal{M}^{\text{s}}_k[0] \gets O_k[0]$}
%  \STATE \hspace*{\algorithmicindent}\parbox[t]{.8\linewidth}{\raggedright   $\bar{P}_k \gets \sum_{l \in \mathcal{M}^{\text{s}}_k[0]} 10\log_{10} \left( \beta_{l,k}[0] \right)$}
% \FOR{$t = 1 \dots T$} 
%     \STATE{$P_k[t] \gets 10\log_{10} \left( \sum_{l \in \mathcal{M}^{\text{s}}_k[t]} \beta_{l,k}[t] \right)$}
%     \IF{$\bar{P}_k - P_k[t] > M^{\text{F}}_{\text{HO}}$}
%         \STATE{$\mathcal{M}_k[t] \gets O_k[t]$}
%         \STATE{$\bar{P}_k \gets 10\log_{10} \left( \sum_{l \in \mathcal{M}^{\text{s}}_k[t]} \beta_{l,k}[t] \right)$}
%         \ELSE
%      \STATE $\mathcal{M}^{\text{s}}_k[t] \gets \mathcal{M}^{\text{s}}_k[t-1]$
%     \ENDIF
% \ENDFOR
% \end{algorithmic}
% \end{algorithm}
By the fixed clustering strategy, 
% \ag{ described in Section xx}. 
% In this case, 
% the UE should trigger a handover if it notices that its DL received 
% power deteriorates too much. \sofie{expected? instantaneous? Or both are the same adduming some kind of channel hardening? } \robbert{expected but they shouldn't differ too much indeed} 
the UE monitors the instantaneous total received power of the cluster $P_k[t]$, via a DL control channel
and triggers a handover if it is significantly below the initial cluster power, i.e.   
$ P_k[t] < \bar{P}_k - M^{\text{F}}_{\text{HO}}$,
%\vida{$P_{k}[t]$ Is the received power from the strongest O-RU? I see that you replaced $P_{k}[t]$ with $max \beta_{l,k}$ in algorithm 2? you could maybe define $P_k[t]$ here?}, 
where $M^{\text{F}}_{\text{HO}}$ is a hysteresis threshold.
The UE then chooses a new primary O-RU and O-DU. The measurement cluster 
is then also updated by the Near-RT RIC. Subsequently, the Near-RT RIC determines a new
serving cluster for that UE.
% \vida{Maybe you could bring it as a footnote that one way to measure the power is by DL pilot proposed in ....}.  \ag{The last sentence bring some confusion, as the paper does not talk about DL pilots.}
The threshold brings a trade-off in the system's performance (higher SE) 
versus 
% \vida{less signaling overhead? I get what you mean but when we say tradeoff, we are comapring two things and mentioning their advantage and selecet one of them based on our preference, no?} 
less signaling overhead (less frequent handovers).
%\sofie{I imply from this that you assume that the UE only has information on the serving set, but when a handover is triggered, the measurement set is determined again?}
%\robbert{Sofie, yes exactly. I added it now}\\
% \vida{So the new primary O-RU must be in the old serving cluster?} 
% \robbert{no, but it should be in the measurement cluster}\vida{Did you mention that in the text?}
% \vida{So RIC always memorizes the measurement cluster of the UEs so in case of handover, 
% to select a new primary O-RU from the cluster. However, what if the UE best O-RU is not in the measurement cluster?
% Does not matter I assume because RIC will assign it to the UE and RIC select the best possible one?}
% \robbert{I made a mistake, the UE selects the best O-RU regardless of measurement or serving cluster, and then signals the RIC to 
% create the measurement cluster around that one} 
 
 % and thus we see this as a valid assumption.  
% It is also possible to transmit the UL measurements back to the UE in a control channel and to use these under 
% mild reciprocity assumptions. 
% We provide pseudocode in Algorithm~\ref{alg:cap}.
% \sofie{how does it learn the statistics? can you explain more precisely?}
% \sofie{which statistics? where are they specified? statistics to one or all serving O-RU?}
\begin{comment}
\begin{figure}
\centering
    \includegraphics[width=0.7\linewidth]{figures/UE-Centric Handover.png}
    \caption{Signalling flow for UE Centric Handover}
    \end{figure}
\end{comment}

%%%%%%%%%%%%%%%%%%%%%%%%%%%%%%
\subsection{Opportunistic Cluster Tracking}
%%%%%%%%%%%%%%%%%%%%%%%%%%%%%%%
%Opportunistic Handover % \vida{Here you are saying that UEs and O-RUs checking for possible handover in two level equally frequently? (both every 1 unit of time) } \robbert{yes}}
We also define a handover strategy for our %opportunistic 
opportunistic clustering.  
% \sofie{It is repetitive and very confusing. You have to say:
% \begin{itemize}
%     \item when a primary O-RU update is triggered. Say simply: only the primary O-RU is guaranteed to be always serving, and the secondary O-RU are determined opportunistically. As a result, the update of the primary O-RU should only depend on the channel towards that primary O-RU. A UE triggers a handover, which is a change of primary O-RU, each time a new .... (explain the lstar and bar tings). 
%     \item when secondary O-RU are opportunistically added (which is not really a handover but only tracking?). 
% \end{itemize}
% s}
In opportunistic tracking, the cluster updates on two different levels: 1) The primary O-RU which is triggered by the UE; 2)  
The opportunistic serving decisions by the O-DUs.
Every UE is connected to one O-RU as its primary one, denoted by $l^{\ast}_k$. 
A change of the primary O-RU requires the UE to perform an actual handover.
% \vida{there was a primary O-RU handover also in fixed clustering (previous subsection) right? I think you can shorten these two sections in only one section and say that when the signal power to the primary O-RU drops below a level, then handover triggers, and a primary O-RU is selected. Then the measurement cluster and serving cluster formation in section 2 follow, right?}. 
% \sofie{was this not said before? }. 
To limit the signaling between the O-DUs and the Near-RT RIC, 
it is logical to make the UE responsible for the handover of its primary O-RU as it is the UE's only 
persistent connection. 
A UE triggers a primary O-RU handover when it detects a significantly higher DL channel gain to 
a different O-RU in its measurement cluster (line 4 in Alg.~\ref{alg:oppHO}). 
% \sofie{I would emphasize here that the metric to decide this can onlly depend on the UE-primary O-RU link quality, as that is the only stable connection in the serving set for the UE in case of opportunistic tracking. This is key and not sure if it is explicitly enough said. Maybe also already say that this metric is then also very sensitive to the UE-primary O-RU link, and could possibly result in more primary O-RU handovers. (in the results section later, it is also not fully clear to me each time what you call a handover for the opportunistic case. I would make the primary handovers, the ones that request a new O-RU and a new meausrement cluster, the costly ones and hence the real handovers.)} 
% \robbert{I now renamed the local decision at the O-RU an \textit{Opportunistic O-RU Reload} to make the distinction more
% clear} 
% \sofie{perfect!}
% \vida{you could just say the UE or the refernce UE instead of new UE?} 
% \robbert{I now changed it to say that that new UE impacts its \textit{new primary O-RU}, I think it's more logical?}
If the new primary O-RU exceeds its limit of served UEs due to that handover, that O-RU drops
the weakest UE it is currently serving opportunistically (line 5-8 in Alg.~\ref{alg:oppHO}).
% \vida{I think the O-RU should drop the weakest UE that it is serving \textcolor{red}{Not as primary O-RU}. But I am thinking what if for all the UEs it is serving, it is THE primary O-RU? then it can not drop any? Or it can? consider the case the for the new O-RU all the currently served users are primary users, and they are all stronger that our refernce UE that want to do handover. Then what new O-RU will do? } 
% \vida{What if this O-RU is already fully loaded by other users?}. 
% \robbert{I should clarify that the limit is for opportunistic serving, not for primary serving}
When the primary O-RU is updated, the measurement cluster changes accordingly, i.e. 
the closest O-RUs to the new primary O-RU become the new 
measurement cluster. % \sofie{as decided by the near-RT RIC?}.
% \vida{same as fixed cluster formation right?}. %\\ % \sofie{THis is not represented in teh algorithm in pseudocode.}\\
% \vida{As here you are defining another type of handover, I would start with "additionally or furthermore" or you can say The O-DU can also .... because this way it shows you are seperating the two handover which are triggered based on somehow different entities}
% \vida{Myabe also mention the names of the handover, no? defined as local O-RU handover...}
% \robbert{Yes, but I'm now thinking to change it to opportunistic O-RU handover.}
% \vida{make sure to put the name after you finalized the name, both are good names}
Additionally, an O-DU dynamically changes the set of UEs it opportunistically serves on its O-RUs with resources not occupied by primary O-RU connections. 
% We call this an \textit{opportunistic O-RU reload}. 
% \vida{here you mean the O-DU in new mesurement cluster to decide if they want to serve the refernce (handover) UE or not? then why do you say UE\textcolor{red}{s}?}.  
% with only minimal communication with the rest of the network. 
% \sofie{But this is not handover, right?}\robbert{depends on how you define handover, I think it's logical to discuss the two levels here, also reworded it now}. 
The O-DUs constantly track the path loss between its O-RUs and UEs in its measurement clusters and 
choose to serve the best ones (line 13-17 Alg.~\ref{alg:oppHO}).
% (via $Q_l^{(w)}[t]$ from Section~\ref{subsec:nettrig}). 
Any changes in the opportunistic serving set do %can happen dynamically as this does 
not change any connections between UEs and their primary O-RU. 
% \sofie{Not super clear? So, UE k had a significantly higher UL channel gain to O-RU l, then all UE with O-RU l in their measurement cluster are doing a new measurement. Then, UE t (can be different than UE k) is served by O-RU l, opportunistically. ? } \robbert{yes, tried to put it more clearly now; see above}.
% \vida{So here maybe some UEs are dropped. Why we need to have this handover? maybe when the UE moves, 
% then it should change the mesaurement cluster at some point so level 1 is not enough. 
% But we didnot have multiple level handover in fixed clustering maybe not need}. 
% \robbert{Vida, yes, here you need to have it at two levels because there is no 
% network-wide controller that can ensure every UE is served by at least one O-RU. Hence, we select
% the UE to take care of it} 
% \sofie{is there a difference between serving as primary O-RU or secondary O-RU? They both count the same, right?}
% \robbert{They count as the same but a UE that wants to use it as a primary should be prioritized over ones that want it as secondary}. 

We introduce the handover threshold as $M^{\text{O}}_{\text{HO}}$ and design it to be the same
for the handover of the primary O-RU which is triggered by the UE and the opportunistic serving of extra UEs % \sofie{addition to what? Confusing. Also, for the UE-triggered handover (primary O-RU reselection) should that not be a relative decision - e.g., handover to new O-RU when a O-RU is found with xdB better gain than the current primary O-RU? That is how handover is done in traditional cellular systems. }.
However, we do acknowledge that it might be interesting to use 
different thresholds for the primary handover and opportunistic O-RU reload % \sofie{what is opportunistic O-RU reload?}
and consider it an important direction for future work. 
% \sofie{OF course! I would expect you to want to avoid pimary O-RU updates, but adding new O-RU opportunistically should be done quite dynamically...?}
% \sofie{But this cluster tracking is not really a handover? Call it maybe just cluster updating?} 
% \sofie{In algorithm 2 the limit was N, there was no difference between primary and secondary O-RU on this limit. 
% Is there a reason why you complicate it here?  } 
%  \robbert{different thresholds for the primary O-RU and opportunistic serving, i.e. keeping a UE that uses an O-RU as primary much longer than one that is added opportunistically}. 
We show the procedure Algorithm~\ref{alg:oppHO}. 
\begin{algorithm}[h]
\caption{Opportunistic Cluster Tracking}\label{alg:oppHO}
\begin{algorithmic}[1]
\STATE{\textbf{At every time $t$}}
\FOR{$k = 1 \dots K$} % \sofie{# for each UE $k$ update if needed the primary O-RU
    \STATE{$\bar{l} \gets \arg\max_l \beta^{\text{dB}}_{l, k}$}
    \IF{$(\beta^{\text{dB}}_{\bar{l}, k}[t] > \beta^{\text{dB}}_{l^{\ast}_k, k}[t] + M^{O}_{HO})  \land (\bar{l} \neq l^{\ast}_k)$}
        \STATE{\textbf{Primary O-RU handover}}
        % \sofie{Define lbar and lstar. Explain also when this main (primary???) O-RU handover is triggered by the UE, and why it is different then for the fixed clustering scheme? }
        % \STATE{$K^{\star}_{l^{\star}_k} \gets K^{\star}_{l^{\star}_k} - 1$}
        \STATE $\mathcal{D}_{l^{\star}}[t] \gets Q_{l^{\star}}^{(N - K^{\star}_l + 1)}[t]$  
        \STATE {$l^{\ast}_k \gets \bar{l}$}
        % \STATE{$K^{\star}_{\bar{l}} \gets K^{\star}_{\bar{l}} - 1$}
        \STATE $\mathcal{D}_{\bar{l}}[t] \gets Q_{\bar{l}}^{(N - K^{\star}_{\bar{l}})}[t]$ 
        % \robbert{is the $\bar{l}$ clearly visibile? I'm also a bit worried that maybe the notation is a bit convoluted?}
        % \sofie{Mention that the measurement set is updated?}
    \ENDIF
\ENDFOR
\FOR{$l = 1\dots L$} % \sofie{For each O-RU $k$ update the set of UE it is serving opportunistically by adding better UE (and dropping worse UE)?}
    \IF{$\exists \bar{k} \notin \mathcal{D}_l[t]: \beta^{dB}_{l, \bar{k}}[t] > \beta^{dB}_{l,k}[t] + M^{O}_{HO}, \quad k \in \mathcal{D}_l[t]$ }
        \STATE{\textbf{Opportunistic O-RU Reload}} %\sofie{I would not call this handover, you defined above that a handover is there when the primary O-RU changes}}
        \STATE{$\mathcal{D}_l[t] \gets Q_l^{(N - K^{\star}_l)}[t]  \cup  \{k: l^{\ast}_k = l\} $}
        \ELSE
         \STATE{$\mathcal{D}_l[t] \gets \mathcal{D}_l[t-1]$}
    \ENDIF
    %  \sofie{I see where you add better UE $\bar{k}$, but I don't see where you drop UE.}\robbert{It's in line 14, only take the best i.e. automatically drop worse ones}
\ENDFOR
\end{algorithmic}
\end{algorithm}
\subsection{Baselines}
By the ubiquitous method, the UE is served by all O-RUs. The UE can move anywhere 
without significant losses because the LSFD combiner 
can calculate new combining weights $\mathbf{a}_k$ based on the changing channel. 
Hence, there is no
handover needed. However, it is helpful to highlight the performance gap to this method. 
% \robbert{I don't think the algorithm is needed here}
% \begin{algorithm}[htb]
% \caption{Cellular Handover}\label{alg:cellHO}
% \begin{algorithmic}[1]
% \FOR{$t=1\dots T$}
% \FOR{$k = 1 \dots K$}
%     \STATE{$\bar{l} \gets \arg\max_{l \notin \mathcal{M}^{s}_k} \beta^{\text{dB}}_{l, k}$}
%     \IF{$(\beta^{\text{dB}}_{\bar{l}, k} > \beta^{\text{dB}}_{l^{\ast}_k, k} + M^{\text{C}}_{\text{HO}})$}
%        \STATE $\mathcal{M}_k^{s}[t] \gets \{\mathcal{L}_c | \bar{l} \in \mathcal{L}_c\}$
%        \ELSE
%        \STATE $\mathcal{M}_k^{s}[t] \gets \mathcal{M}_k^{s}[t-1]$
%     \ENDIF
% \ENDFOR
% \ENDFOR
% \end{algorithmic}
% \end{algorithm}
By the cellular method, a UE requests a handover when it detects that it has a significantly higher 
 DL channel gain to an O-RU in a different O-DU than the one by which it is currently served based
 on a hysteresis threshold.
\section{Control and Data Plane}
In this section, we give a brief comparison of the signalling costs associated with the 
different methods. 
% with the methods above in Table~\ref{tab:costs}. The costs for UL processing are taken from \cite{SIG-109} and expressed in floating point operations. We then link these to our clustering methods. We first list the cost of these methods here per UE - O-RU association. UL channel estimation costs $C_{\text{Ch. Est}}(|\mathcal{D}_l|) = \left((3N^2 + N)\frac{K}{2} + (N^3 - N)\frac{1}{3} \right)$\vida{$N^3-N/2$ is for matrix inversion? $3N^2/2$ is for multiplication of the inverse matrix with R and y, following formula 4? what is N for? Also, why we have K here? shouldn't it be $\mathcal{D}_l$?}, calculating the local decoder costs $C_{\text{L-MMSE}}(|\mathcal{D}_l|) = \left((3N^2 + N)\frac{|\mathcal{D}_l|}{2} + (N^3 - N)\frac{1}{3} \right)$ and finding the LSFD decoder costs $C_{\text{LSFD}}(|\mathcal{M}_k^s|) = |\mathcal{M}_k^s|^2 + \frac{|\mathcal{M}_k^s|^3 - |\mathcal{M}_k^s|}{3}$\vida{LSFD coefficient calculation is similar to combiner only the $N\times N$ matrix is replaced by $|\mathcal{M}_k^s|\times |\mathcal{M}_k^s|$ matrix and $N\times 1$ vector is replaced with $|\mathcal{M}_k^s|\times 1$. Then why the $C_{LSFD}$ calculation is different than $C_{L-MMSE}$?}. 
% Finally, the signaling 
% of statistics and samples is upper bounded by the worst case where all serving O-RUs, except the primary O-RU, belong to different O-DUs than the primary O-DU.
For UL data plane processing, the values are derived from \cite{SIG-109}. O-RU $l$ needs to signal $\tau_u$, which is the number of data symbols in a coherence block, samples %$\hat{s}_k^{l,k}$
 to the O-DU for every user $k$ in $\mathcal{D}_l$. These are 
then combined in the O-DU to $\tau_u$ samples of $\hat{s}_k^{c}$ for every unique user served by the 
O-DU. For every UE $k$ O-DU $c$ is serving as a primary O-DU, O-DU $c$ receives $\tau_u$ samples for every unique O-DU that controls the O-RUs in the set $\mathcal{M}_k^s \setminus (\mathcal{M}_k^s \cap \mathcal{L}_c)$.
Additionally, for every UE $k$ they 
are serving but not as primary O-DU, O-DUs transmit $\tau_u$ samples $\hat{s}_k^{c}$ 

For cluster formation, the costs vary significantly 
for the two methods. For the fixed method, every time the 
UE triggers a handover, the O-DUs in the measurement cluster must transmit $|\mathcal{M}_k^m|$ UL path loss  $\beta_{l,k}$ measurements to the Near-RT RIC.
So in the worst case, if every user triggers a handover at the same time, 
the traffic to the Near-RT RIC reaches $|\mathcal{M}_k^m|K$. The opportunistic 
method only requires the new primary O-RUs to notify the Near-RT RIC of its UEs. 
% \sofie{So in the worst case, if every user triggers a handover at the same time, 
% the traffic to the Near-RT RIC reaches XXX. }
% between O-DUs where $\tau_u$ is the number of data samples per coherence block. 
% Depending on how many O-RUs of $\mathcal{M}_k^s$ are on the primary O-DU this becomes 
% less tight. 

% An O-DU then needs to recieve this many samples per UE it is serving as 
% a primary and Additionally, we need 
% to transfer $F_{\text{statistics}}(\{|\mathcal{D}_l|\}_l,\mathcal{M}_k^s) = \sum_{l \in \mathcal{M}_k^s}|\mathcal{D}_l|\frac{3|\mathcal{S}_k| + 1}{2}$\vida{I did not understand here, so if O-RU $l$ is serving $\mathcal{D}_l$ users, then it sends the inner product of each users combining vector with others channel ($\mathcal{D}_l\times \mathcal{D}_l$) scalars to the primary ODU no? and second how you define $\mathcal{S}_k$? the users that have most interference on user k? isn't it all the users that share at least one serving O-RU with user k, so $\mathcal{S}_k=\cup_{l\in\mathcal{M}_k^s} \mathcal{D}_l$ ? Also, why we have 3/2 as a factor? and +1? maybe +1 is for user k channel gain $\mathbb{E}\{g_{kl}^*g_{kl}\}$? plus, why I can not find $F_{\text{statistics}}(\{|\mathcal{D}_l|\}_l,\mathcal{M}_k^s)$ in the table? I mean the exact notation as in the text? Also, your summation is on serving O-RU of only user k and you did not mention that it is per user } every time new LSFD coefficients are calculated. 
% For the fixed strategy and the opportunistic strategy, respectively $|\mathcal{M}_k^s|$ and $|\mathcal{D}_l|$ can be tuned

\section{Simulation and Numerical Results}
\label{sec:results}
%Finally, w
We  
quantify the performance of the different clustering and handover methods.
% For the figures that relate to the clustering, we provide the canonical, ubiquitous cell-free case as an upper bound 
% for the performance of our clustering/handover algorithms; because all O-RUs cooperate in this scenario.
We provide the ubiquitous case as an upper bound and the cellular case as a lower bound on the performance; for this case, we use a hysteresis 
parameter of 2 dB. The parameters for the simulation are listed in Table~\ref{table:sim}.
The O-RUs are located in a square grid, this grid is divided into uniformly sized squares for each O-DU, 
the O-RUs for a specific O-DU are then placed randomly in the subsquare for its O-DU via a uniform
distribution. 
% \ag{Is there any figure that you can refer/cite?} 
The UEs are then also randomly placed in the grid according to a uniform distribution.
To ensure consistent performance across the simulation area, 
we replicated the initial setup eight times, resulting in a wrap-around scenario with a 3 × 3 tile configuration. %This way, the UEs at the edge of the initial grid are  
% The UEs move in a straight line over a time period of 10 seconds. 
\begin{table}[htb]
\centering
\begin{tabular}{l|r|l|r}
\textbf{Parameter}     & \textbf{Value} & \textbf{Parameter}     & \textbf{Value}\\ \hline \hline
K                     &  40    &  L                      & 36   \\     \hline  
N                      & 4   & C                      & 9   \\ \hline
 $\sigma^2_{\text{ul}}$ & -94 dBm &  $\tau_p$               & 100   \\\hline
 Grid Size              & 1 x 1 km &  $T_s$                  & 0.5s    \\  \hline
Number of Setups       & 25     &   $\xi$                   & 10°   \\     \hline
   $|\mathcal{M}^{\text{s}}_k|$              &    16      & Simulation Time  & 10s   \\  \hline \hline
\end{tabular}
\caption{Simulation Parameters 
% \robbert{Is it more clear if I add the numbers for the initial tile
% here or the numbers for the complete wrap around grid? Currently it's the initial grid}}
}
\label{table:sim}
\end{table}%

\subsection{Handover Frequency}
\begin{figure}[tb]
\centering
\begin{tikzpicture}
\begin{axis}[ybar,
        width=0.8\linewidth,
        ymin=0,
        ymax=1,  
        % ymode=log,
        % xtick = data
        ylabel={Handover Frequency [$s^{-1}$]},
        xlabel={UE Speed [km/h]},
        xtick = {1,2,3,4},
            grid=major, % Display a grid
          grid style={dashed,gray!30}, % Set the style
        xticklabels = {
            3,
            30,
            60,
            120,
        },
        bar width=4pt,
        major x tick style = {opacity=100},
        enlarge x limits=0.15,
        % minor x tick num = 1,
        % minor tick length=2ex,
        legend pos = north west,
        legend style={nodes={scale=0.6, transform shape}},
        ]
        \addplot[fill=blue]  table[x=speed, y=2, col sep=comma] {figures/resultsFinal/UECentricHandover.csv}; %
                \addlegendentry{Fixed (2 dB)}
    \addplot[fill=red]  table[x=speed,y=3, col sep=comma] {figures/resultsFinal/UECentricHandover.csv};
            \addlegendentry{Fixed (3 dB)}
    \addplot[fill=brown]  table[x=speed,y=2, col sep=comma]  {figures/resultsFinal/APCentricHandover.csv};
            \addlegendentry{Opp. (2 dB)}
        \addplot[fill=black]  table[x=speed,y=3, col sep=comma]  {figures/resultsFinal/APCentricHandover.csv};
                \addlegendentry{Opp. (3 dB) }
        \end{axis}
\end{tikzpicture}
\caption{Handover frequency for the different methods for two different thresholds 
($M^{\text{O}}_{\text{HO}}$/$M^{\text{F}}_{\text{HO}}$) and varying UE's speeds.}% \ag{Small figure}}
\label{fig:handovers}
\end{figure}
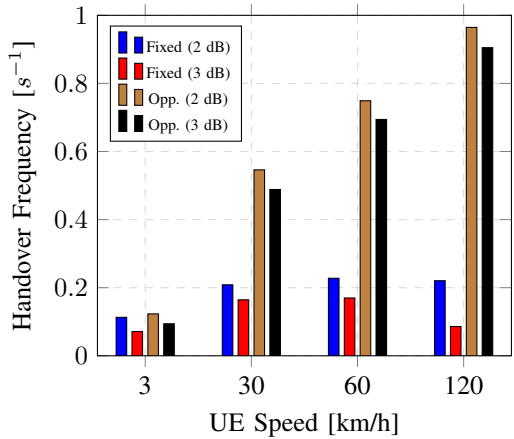
Figure~\ref{fig:handovers} quantifies the handover frequency for different speeds and 
different threshold values ($M^{\text{O}}_{\text{HO}}$ and $M^{\text{F}}_{\text{HO}}$). 
Fixed clustering, which is expensive for the Near-RT RIC, 
exhibits a lower handover frequency than opportunistic clustering for the same threshold. 
This is attributed to the tracking of cluster-wide channel gain which 
is less sensitive to the UE's mobility than the channel gain for only the primary O-RU. 
For both methods, the number of handovers increases with speed and decreases with a rising threshold. This 
highlights a vital trade-off, a high frequency of handovers improves the SE significantly, 
especially for high speeds. This becomes expensive for fixed clustering
% \ag{Was E2 mentioned before? If not, include at the beginning of section II} 
as the Near-RT RIC
becomes very loaded with handover procedures whereas the opportunistic cluster tracking only
requires minor signaling. %\sofie{not every handover, as you have two types of handover. Make sure it is clear that the opportunistic handover frequency is the primary handover one. Is it? confused...} \robbert{What I want to say is that the handover can be done per O-DU, it is only the primary that 
% needs real change. So I argue that even doing a primary handover is much cheaper}.  

\subsection{Mean Spectral Efficiencies}
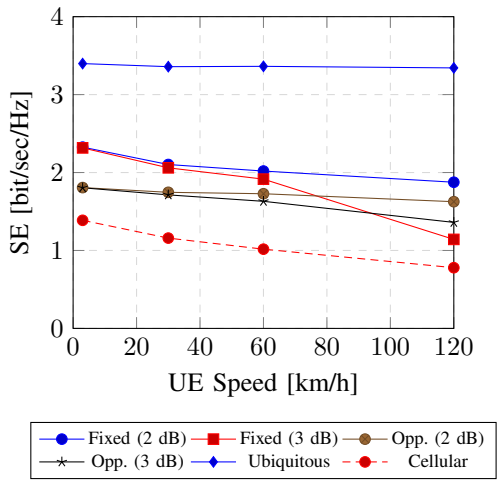
\begin{figure}[htb]
\centering
  \begin{tikzpicture}
      \begin{axis}[width=0.75\linewidth, % Scale the plot to \linewidth
          grid=major, % Display a grid
          grid style={dashed,gray!30}, % Set the style
          ylabel={SE [bit/sec/Hz]}, % Set the labels
          xlabel={UE Speed [km/h]},
          xmin=0,
          xmax=120,
          ymin=0,
          ymax=4, 
          mark repeat=1,
          legend columns=3,
          %y unit=\si{\ampere},
          legend style={nodes={scale=0.7, transform shape}},
          legend style={at={(0.5,-0.3)},anchor=north}, % Put the legend below the plot
          x tick label style={rotate=0,anchor=north} % Display labels sidewayn
        ]
        \addplot table[x=speed, y=2, col sep=comma,mark repeat=10] {figures/resultsFinal/UECentricSpeeds.csv}; 
        \addlegendentry{Fixed (2 dB)}
            
        \addplot  table[x=speed, y=3, col sep=comma] {figures/resultsFinal/UECentricSpeeds.csv};
        \addlegendentry{ Fixed (3 dB)}

        \addplot  table[x=speed, y=2, col sep=comma] {figures/resultsFinal/APCentricSpeeds.csv};
        \addlegendentry{Opp. (2 dB) }

        \addplot table[x=speed, y=3, col sep=comma] {figures/resultsFinal/APCentricSpeeds.csv};
        \addlegendentry{Opp. (3 dB)}
        
        \addplot table[x=speed, y=upper, col sep=comma] {figures/resultsFinal/UECentricSpeeds.csv};
        \addlegendentry{Ubiquitous}
        \addplot table[x=speed, y=lower, col sep=comma] {figures/resultsFinal/UECentricSpeeds.csv};
        \addlegendentry{Cellular}
      \end{axis}
    \end{tikzpicture}
    \caption{Mean SE for different UE speeds}% \ag{Small figure}}
     % \adam{Is the "O-RUs serving a user" the same as the serving cluster size?}}\robbert{yes}
    % \label{fig:results7}
    \label{fig:comparison}
\end{figure}

Figure~\ref{fig:comparison} illustrates the influence of the threshold on the mean spectral efficiency (SE) as a function of UE speed. 
For the fixed clustering, the mean SE drops 
with increasing UE speed and even moreso if the HO threshold is chosen too large.
In our setup, performance then degrades beyond that of the opportunistic method.
% \sofie{why is fixed better for low speed? they have different clusters? serving threshold too low as you fix it to the handover threshold? Maybe mention that the opportunistic perofmrnace could easily be scaled to have higher SE. Do you by the way track the average number of serving O-RU? then you would see that the cost of the fixed is also higher in the data plane, right? (always N serving)?}
% \vida{and based on a different criteria, namely single O-RU right? maybe you dont need to mention that handover thereshold of cellular is low, until beyond that of cellular cae seems enough.}. 
The performance of the opportunistic clustering %seems to be influenced much less
%by the actual threshold that is taken
is less sensitive to the threshold value
than the fixed clustering
% \vida{than the fix clustering. because your next sentence is about fixed clustering so maybe make sense to mention it in this sentence}. %UE-triggered handover however is much more sensitive to the 
%chosen threshold. 
due to the UE only triggering handover if the total received power 
decreases too much in the fixed clustering. %if 
The opportunistic clustering allows a new O-RU to easily start serving a new UE
when that O-RU detects it as sufficiently strong. 
% \vida{Does it? Actually this is one of my question in the handover section of the opportunistic handover}
% \vida{I think opportuistic clustering is better because the handover happens based on the received power from single O-RU but for fixed clustering, 
% we tolerate more by taking the total received power into account} 
Hence, the opportunistic strategy provides more 
fine-grained control. This does, however, come at the cost of not having a deterministic number of 
O-RUs that serve a particular UE as it is impossible for our algorithm
%\sofie{I don't agree that it is impossible. in your algorithm it is just not the case. it is a trade-off between network wide QoS control and cost....} 
to guarantee any quality of service 
beyond that provided by the primary O-RU. 
% \sofie{ok, so report on the O-RU number variation? Is because of that the SE variation also higher for the opportunistic one, which the fixed strategy has a more stable SE? A lot of things to analyse deeper for the journal!}. 
% \vida{why is it a cost? you mean maybe some UE got drop off completely, no O-RU to serve them?}. 
% \vida{Not sure if also can officially start a sentence, maybe use: Furthermore, it is observed that the ubiquitous case is barely..} 
Furthermore, the ubiquitous case is barely affected 
by increasing speed of the UE. 

\section{Conclusion}

In this work, 
% we mapped an UL detection method from the CF mMIMO state-of-the-art to the O-RAN architecture. 
we have introduced a novel temporal channel for correlated Rayleigh fading in CF mMIMO networks. 
% \ag{Do we?}\robbert{Yes, section 3e}
We introduced two clustering and handover strategies, mapped them to the O-RAN architecture and benchmarked them via our selected UL detection method based on the mean SE. We find that the opportunistic clustering provides more 
fine-grained control than the fixed clustering strategy in the presence of UE mobility.
Furthermore, we have demonstrated that CF mMIMO is much more resilient against UE mobility compared to
classical cellular systems.
\bibliographystyle{IEEEtran}
\bibliography{references.bib}
\end{document}